%% file: make_all.tex
\documentclass{he_symp}
\usepackage{psfig,graphicx,epsfig}
\usepackage{color}
\usepackage{amsmath,amssymb,epic,eepic,array}
\begin{document}
\renewcommand{\FirstPageOfPaper }{ 87}\renewcommand{\LastPageOfPaper }{ 90}\include{he_symp_nicastro}
\clearpage

\end{document}

%% file: he_symp_nicastro.tex
\def\P{PSR~B1937+21}
\def\B{{\it Beppo}SAX}
\newcommand{\fu}{erg~cm$^{-2}$~s$^{-1}$}
\def\fdeg{\hbox{$\,.\!\!^{\circ}$}}
\def\loe{\lower 0.6ex\hbox{${}\stackrel{<}{\sim}{}$}}
\def\goe{\lower 0.6ex\hbox{${}\stackrel{>}{\sim}{}$}}

\title{The 2-10 keV emission properties of PSR B1937+21}
\author{L. Nicastro\inst{1}
 \and G. Cusumano\inst{1}
 \and L. Kuiper\inst{3}
 \and W. Becker\inst{2}
 \and W. Hermsen\inst{3}
 \and M. Kramer\inst{4}
}
\institute{IASF--CNR, Via U. La Malfa 153, 90146 Palermo, Italy
\and Max--Planck--Institut f\"ur extraterrestrische Physik,
 Giessenbachstra{\ss}e, 85740 Garching, Germany
\and  SRON National Institute for Space Research, Sorbonnelaan 2, 3584 CA Utrecht, The Netherlands
\and  University of Manchester, Jodrell Bank, Macclesfield SK11 9DL,
      United Kingdom
}
\maketitle

\begin{abstract}
We present the results of a \B\ observation of the fastest pulsar known: \P.
 The $\sim 200$ ks observation (78.5 (34) ks MECS (LECS) exposure times) 
 allowed us to investigate with high
 statistical significance both the spectral properties and the pulse
 profile shape.
 The absorbed power law spectral model gave a photon index of $\sim 1.7$ and
 $N_{\rm H} \sim 2.3\times 10^{22}$ cm$^{-2}$. These values explain both
 {\it a.} the ROSAT
 non-detection and {\it b.} the deviant estimate of a photon index of $\sim 0.8$
 obtained by ASCA. The pulse profile appears, for the first time,
 clearly double peaked with the main component much stronger than the other.
 The statistical significance is $10 \sigma$ (main peak) and $5 \sigma$
 (secondary peak).
 The 1.6--10 keV pulsed fraction is consistent with 100\%;
 only in the 1.6--4 keV band there is a
 $\sim$ 2$\sigma$ indication for a DC component. The secondary
 peak is detected significantly only for energies above $3\div 4$ keV.
 The unabsorbed
 (2--10 keV) flux is $F_{2-10} = 3.7\times 10^{-13}$ \fu, implying a
 luminosity of $L_X = 4.6\times 10^{31} \, \Theta$ ($d/3.6$ kpc)$^2$
 erg s$^{-1}$ and an X-ray efficiency of $\eta = 4\times 10^{-5}\, \Theta$,
 where $\Theta$ is the solid angle spanned by the emission beam.
 These results are in agreement with those obtained by ASCA.
\end{abstract}

\section{Introduction}

The various X-ray satellites, from Rosat to Chandra, observed about 10
 millisecond pulsars (MSPs) (see e.g. Becker 2001),
 but good spectral and temporal informations exist only for half of them.
 In fact MSP X-ray observations are usually affected not only by low
 statistics but also by insufficient time accuracies to perform detailed
 periodicity and timing analysis.
 X-ray observations demonstrate that the X-ray emission from MSPs is
 generally not of
 thermal origin, and those with the hardest spectra appear to be
 objects with strong magnetic fields at the light-cylinder
 $B_L$ ($R_L = cP/2\pi$; see Saito et al. 1997,
 Kuiper et al. 1998, Kuiper et al. 2000, Becker 2001).
 Given the correlation between spin-down power and X-ray luminosity
 (see Verbunt et al. 1996, Becker and Tr\"umper 1997 and
 Takahashi et al. 2001), another
 important quantity is the spin-down flux density at Earth ($\dot{E}/4\pi d^2$).

\P, with a period $P = 1.56$ ms, was the first and still the fastest MSP known.
 In spite of its low surface magnetic field strength of
 $B_S = 4.1\times 10^8$ G, its magnetic field at the light-cylinder
 is the highest of all known pulsars: $B_L \simeq 1\times 10^6$ G,
 very similar to that of the Crab pulsar.
 Its spin-down flux density is $\sim 2\times 10^{-10}$ \fu, which is 3 times
 higher than for PSR J0218+4232 but 3 times lower than for PSR B1821$-$24.

Here we report on the results obtained by a \B\ observation in the energy
range 1.6--10 keV.

\section{Observation and spatial analysis \label{obssec}}
\B\ observed \P\ on May $1^{\rm st}$ 2001.
 The $\sim 200$ ks observation time resulted in a total effective on-source
 time of $\sim 78.5$ ks ($\sim 34$ ks) with the MECS (LECS).
 To extract the source photons from LECS (0.1--10 keV) and
 MECS (1.6--10 keV), we used:
\begin{itemize}
\item a standard spatial analysis with optimized extraction radius and
 energy range;
\item a Maximum Likelihood (ML) approach to extract the number of counts
 assigned to the pulsar taking into account the presence of other sources
 and the background simultaneously.
\end{itemize}
In the first case we estimated the local background (in a circular region
 in the field of view $10'$ away from the pulsar) and compared the
 results with the standard one from archival
 blank sky usually used in the spectral fitting. We found that the local
 background is 50\% and 15\% higher for LECS and MECS respectively.
 The extraction radii which optimized the signal to noise ratio were
 $3'$ (LECS) and $2'$ (MECS) resulting in a collection of 106 (0.5--8 keV)
 and 369 (1.6--10 keV) photons.
 In spite of the lower exposure time and sensitivity of LECS we were able
 to detect a significant signal in the 0.5--8 keV range for this instrument.
 Inclusion of these softer photons in the spectral fit allowed us to better
 constrain the hydrogen column density toward the pulsar ($N_{\rm H}$).
Using the ML approach we could extract the source photons modeling the
 contribution to the total counts from additional field sources and the
 (assumed flat) background. This method was successfully used in several
 other cases of weak source (see e.g. Kuiper et al. 1998, Mineo et al. 2000).

\section{Temporal analysis}
\begin{figure}
\centerline{\psfig{file=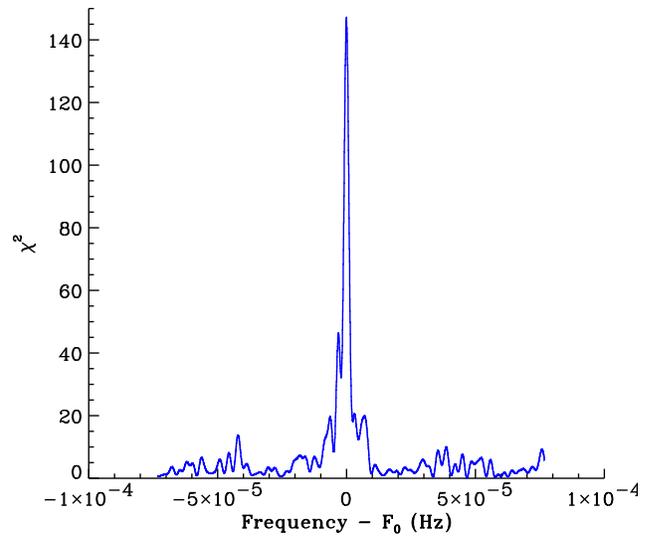,width=8.8cm,clip=} }
\caption{
 The $\chi^2$ significance as function of trial folding frequency.
 The 0 is aligned with the maximum.
\label{chi2}}
\end{figure}

The arrival times of the extracted photons were converted to the Solar System
Barycentric (SSB) Frame using the BARYCONV\footnote{see
http://www.sdc.asi.it/software/saxdas/baryconv.html} code and then searched
for periodicities by folding them with trial frequencies around the
radio one (see updated radio ephemeris in Table \ref{tabeph}).
The $\chi^2$ value plotted versus trial frequency is shown in
Fig. \ref{chi2}.
The distribution had a clear maximum, but we found the corresponding
pulse frequency deviating from the radio value by $\simeq -7.6 \times 10^{-6}$
Hz ($f_{R} - f_{X}$).
This is about ten times the statistical error of $\sim 8 \times 10^{-7}$ Hz.
To investigate this discrepancy we first verified the correctness of the
SSB conversion code by also using independent software.
We verified that all the routines involved in the processing were
working properly.
We then verified the OBT (On-Board Time) to UTC time conversion routine.
To this aim we:
\begin{enumerate}
 \item analyzed separately two sections of the
 observation; the results confirmed those of the full observation;
 \item used different polynomial order for the OBT--UTC fit and various
 high deviation points (de-)selection criteria;
 in all cases the results were the same;
 \item verified the (X-ray) pulse frequency of the Crab
 during the observation performed on April 2000 and on September 2001;
 these observations are shorter than the \P\ one (in particular the
 2001 observation), nevertheless
 in the first case we found a discrepancy which is outside the
 statistical errors.
\end{enumerate}
We concluded that the difference had likely
to be attributed to incorrect photon time markers.
We investigated the possible reasons of that with the \B\ mission
director. The answer was that, excluding an unlikely (and unobserved)
UTC clock drift (which comes out of the Malindi ground station GPS
clock), the only likely explanation could be an on-board
electronics problem. In particular it could be an internal drift of
the clock which tags the ``base band'' packets. Performance degradation
with time of the clock is not unexpected and usually not considered critical
for detectors not specifically built for high precision timing.
Also, as we verified with the Crab, the problem is less
severe (or not visible) for pulsar with relatively long periods or
for relatively short observations.
Even though other (we believe more unlikely) explanations could be found,
we decided to correct the (possible) clock drift by estimating the
required additional slope to add in the OBT--UTC conversion.
The result was a constant slope of $\sim 1\times 10^{-8}$ s s$^{-1}$.
We verified this value (actually a value slightly lower) could also correct
the 2000 Crab observation ($\sim 70$ ks elapsed time) while it had no
effect on the 2001 observation ($\sim 24$ ks elapsed time).
\begin{table}
\caption{Ephemeris of PSR B1937+21 (from observations performed at Effelsberg).
 \label{tabeph}}
\begin{flushleft}
\begin{tabular}{ll}
\hline
\noalign{\smallskip}
{\bf Parameter} &  {\bf Value}  \\
\hline
\noalign{\smallskip}
Right Ascension (J2000) &  19$^{\rm h}$ 39$^{\rm m}$ 38\fs560    \\
Declination (J2000) &  21$^\circ$ $34'$ 59\farcs14     \\
Epoch validity start/end (MJD) &  51639 -- 52047 \\
Frequency &  641.928246349481 Hz\\
Frequency derivative &  $-4.331\times 10^{-14}$ Hz s$^{-1}$ \\
Epoch of the period (MJD) &  51843.0 \\
RA proper motion &  $-0.15$ mas yr$^{-1}$ \\
Dec proper motion &  $-0.47$ mas yr$^{-1}$ \\
\noalign{\smallskip}
\hline
\end{tabular}
\end{flushleft}
\end{table}

\begin{figure}
\centerline{\psfig{file=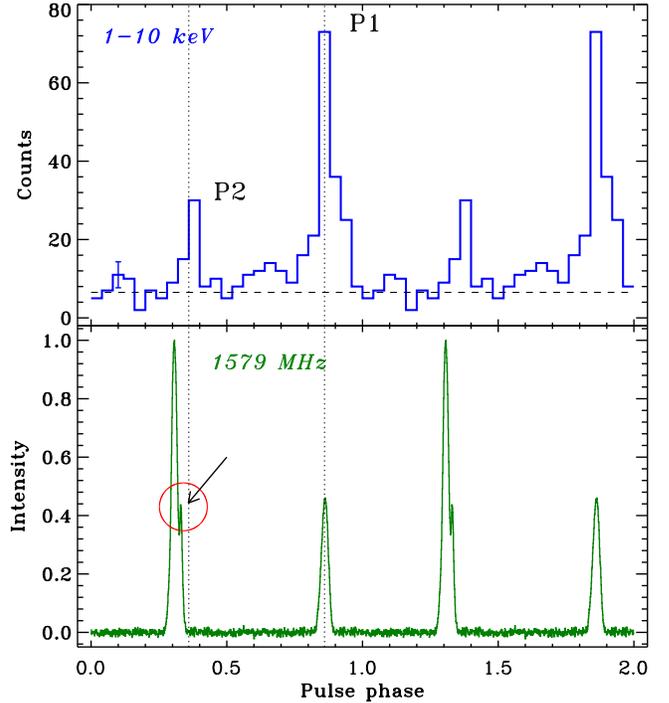,width=8.8cm,clip=} }
\caption{
 The 1.6--10 keV X-ray profile and the radio one arbitrarily aligned
 using the main X-ray peak. The two vertical dotted lines are 0.5 apart.
 The X-ray phase separation is 0.48 like for the radio but 0.5 shifted
 in phase (if the ASCA result is correct). In this case we can also
 confidently exclude that the secondary X-ray peak is associated with
 the feature visible in the main radio peak (which instead would mimic
 the X-ray feature in P1).
\label{profxr}}
\end{figure}

\begin{figure}
\centerline{\psfig{file=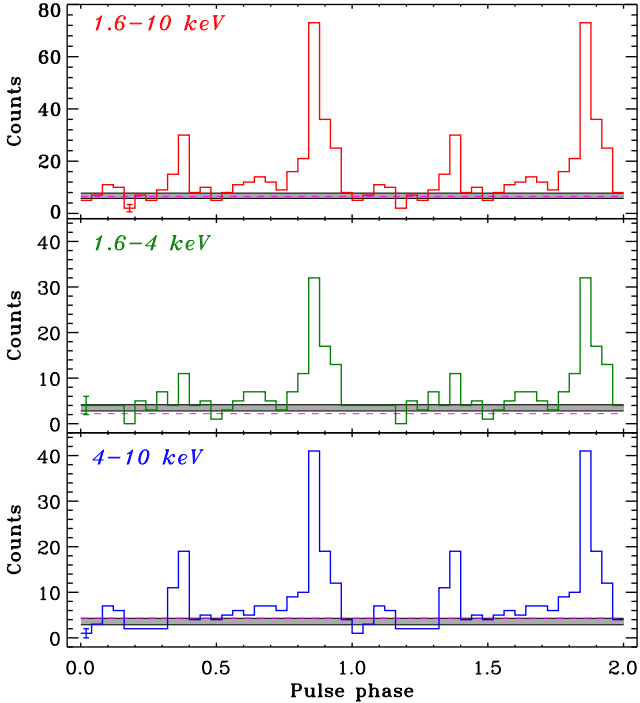,width=8.8cm,clip=} }
\caption{
 The X-ray pulse profile in the full MECS X-ray band
 1.6--10 keV and two sub-bands. The shaded areas show the estimated
 DC level ($\pm 1 \sigma$), while the dashed lines indicate the measured
 background level. In the 1.6--4 keV energy band the secondary peak
 is detected at $\loe 3 \sigma$ level and is compatible with the DC level.
 In the 4--10 keV band instead the detection level is $4 \sigma$
 (above background or DC level).
\label{prof3x}}
\end{figure}

Counting statistics allowed us to perform temporal analysis only for the
MECS data (actually a search on the LECS photons didn't reveal any
significant modulation).
 As already mentioned, a periodicity search analysis
 was performed on the 369 photons (1.6--10 keV) around the nominal radio
 period, after converting the arrival times to the SSB.
 Figure \ref{profxr} shows the folded light curve together with the radio one
 (from the EPN archive\footnote{see http://www.mpifr-bonn.mpg.de/pulsar/dat/}
 -- Kramer et al. 1998). The primary peak (P1) is detected at $10 \sigma$
 level and, for the first time, we also detect the secondary peak
 (P2 -- significance $\sim 5 \sigma$) and estimate a relative phase separation
 of $0.48\pm0.04$. Data collected by ASCA allowed Takahashi et al. (2001)
 to claim an alignment of the main X-ray peak with the radio interpulse.
 We then arbitrarily aligned the two profiles (the BSAX on-board clock
 absolute calibration does not allow to do it for a MSP).
 The vertical dashed lines in the figure are exactly 0.5 apart.
 The radio peaks separation is instead 0.4794(8) so it would be
 tempting to consider the ASCA result incorrect and consider P1 be coincident
 with the main radio peak. The X-ray peak widths are compatible with the
 instrument time resolution limit ($\sim 100$ $\mu$s).
Figure \ref{prof3x} shows the X-ray pulse profile in 3 energy ranges.
 In an attempt to quantify the evidence for a pulsar DC component,
 we applied the bootstrap method proposed by Swanepoel et al. (1996),
 which allows us to estimate the DC level (sky background
 and possibly a source component) in the pulse profile.
 We can then determine the DC and pulsed fractions of the pulsar by
 taking the background level from the spatial analysis into account.
 However, in the available form, the bootstrap method is able to find only one
 unpulsed interval and cannot account for systematic errors.
 One should therefore realize that the quoted errors on the calculated
 parameters are only statistical. Note in Fig. \ref{prof3x} how the DC level
 is compatible with the background level. We find that over the total
 energy range 1.6--10 keV the source is consistent with being 100\% pulsed;
 the $2\sigma$ lower limits for the pulsed fraction in the energy
 ranges 1.6--10, 1.6--4 and 4--10 keV are 88\%, 66\% and 95\%, respectively.
 Only for the lower 1.6--4 keV energy interval we find 
 an indication for a DC component of $(18.7 \pm 7.7)$\%.
 An increase of P2 compared to P1 with energy is clearly visible.
  
\section{Spectral analysis}
\begin{table}
\caption{Power law fit parameters using photons extracted with the
 standard and the Maximum Likelihood methods.
\label{spetab}}
\begin{tabular}{lcccc}
\hline
\noalign{\smallskip}
 {\bf Method} & \boldmath$K$\unboldmath$^{\rm a}$ & \boldmath$\alpha$ &
 \boldmath$N_{\rm H}$ & \boldmath$F_{2-10}$\unboldmath$^{\rm b}$ \\
 &  &  &  ($10^{22}$ cm$^{-2}$) & \\
\hline
\noalign{\smallskip}
{\sl STD} & 0.85 & $1.66^{+0.07}_{-0.06}$ &
  $2.27^{+0.58}_{-0.47}$ & $3.7\pm 0.4$  \\
\noalign{\smallskip}
{\sl ML} & $1.12^{+0.10}_{-0.10}$ & $1.71^{+0.06}_{-0.08}$ &
  $2.10^{+0.56}_{-0.42}$ & $3.7 \pm 0.4$ \\
\noalign{\smallskip}
\hline
\end{tabular}
\begin{list}{}{}
\item Note: all quoted uncertainties are $1\sigma$ confidence.
 It is $N_{ph}(E) = K\times E^{-\alpha}$.
\item[$^{\rm a}$] Normalization at 1 keV ($10^4$ ph s$^{-1}$ cm$^{-2}$).
\item[$^{\rm b}$] Unabsorbed flux in 2--10 keV ($10^{-13}$ \fu).
\end{list}
\end{table}

The spectral analysis was performed using photons extracted in two different
 ways: {\em a.} those in a (S/N optimized) circular region centered on the
 pulsar position (standard method); {\em b.} those from the ML analysis,
 which consider simultaneously the source, the background and a number of
 field sources.
 In both cases the best fitting model was an absorbed power law. A Black
 Body model did not fit the data satisfactorily.
 The two methods gave consistent results
 and are summarized in Tab. \ref{spetab}.
 The absorbed flux (2--10 keV) is $3.1\times 10^{-13}$ \fu.

%
\begin{figure}
\centerline{\psfig{file=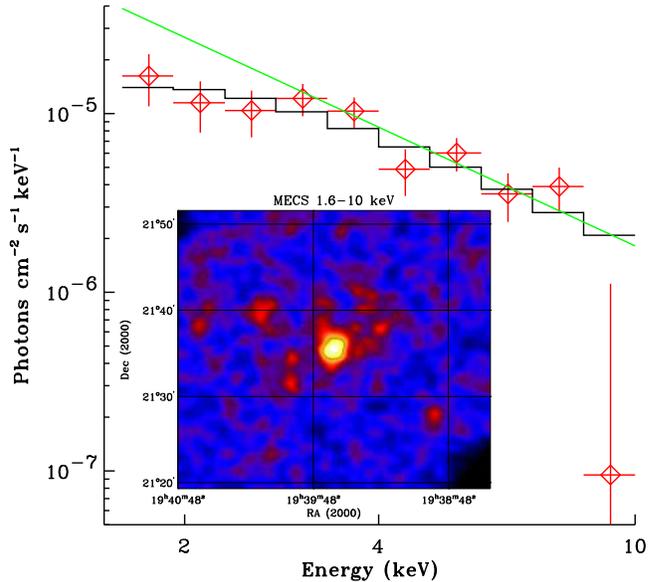,width=8.8cm,clip=} }
\caption{
 The 1.6--10 keV spectrum and best fit model with the unabsorbed power
 law ($\alpha = 1.66$). It is evident the strong absorption below $\sim 3$ keV.
 In the inset the smoothed MECS image around the PSR position is shown.
\label{speima}}
\end{figure}

The derived (unabsorbed) luminosity is $L_X = 4.6\times 10^{31} \, \Theta$
 ($d/3.6$ kpc)$^2$ erg s$^{-1}$ while the X-ray conversion efficiency is
 $\eta = L_X/\dot{E} = 4\times 10^{-5}\, \Theta$,
 where $\Theta$ is the solid angle spanned by the emission beam.
Note that the column density is 10 times greater than the radio DM derived
 one (by assuming 10 H atoms for each e$^-$) and is only
 marginally consistent (within $2 \sigma$) with the Galactic value
 $1.4\times 10^{22}$ cm$^{-2}$.
 In addition the photon index is significantly softer than that found for
 PSR J0218+4232 ($0.94 \pm 0.22$) and PSR B1821$-$24
 ($\sim 1.1$ using RXTE data,  Kawai and Saito 1999).
 Also for the latter pulsar (placed in the globular cluster M28, at a
 distance of $\sim 5$ kpc, galactic latitude $-5\fdeg6$) its $N_{\rm H}$
 $\simeq 3\times 10^{21}$ cm$^{-2}$, consistent both with the
 Galactic and DM derived one.
 Being the line of sight to \P\ tangent to the Galactic spiral arms
 ($l \simeq 57.51$, $b \simeq -0.29$, see Taylor and Cordes 1993)
 one can invoke a particularly low e$^-$/$N_{\rm H}$ ratio in that
 direction due to the lack of ionizing sources.
 However, using the relation $N_{\rm H}=1.79 \times 10^{21} A_V$ of Predehl
 and Schmitt (1995), Verbunt at al. (1996) set an upper limit of
 $2.2 \times 10^{21}$ cm$^{-2}$ for the column density, i.e. in agreement
 with the DM value.
 More simply, the measured high $N_{\rm H}$ could suggest for \P\ a
 true distance greater than the 3.6 kpc derived from the
 Taylor and Cordes (1993) model. If instead this value is correct
 then the continuing radio timing measurements
 should be able to detect its effect on the timing fit residuals very soon.
 We also note that the fit parameters are consistent with the
 non-detection in the ROSAT HRI energy range (0.1--2.4 keV).

\section{Conclusions}

We detected the double peak profile of the fastest pulsar known.
The 1.6--10 keV pulsed fraction is consistent with 100\%, only 
for the band 1.6--4 keV there is an indication for a DC component
(18.7 $\pm$ 7.7 \%).
The secondary (X-ray) peak is significantly detected only above 3--4 keV
and the ratio primary/secondary decreases with energy.
It was not possible to perform an absolute phase comparison
with the radio profile, but, {\em purely} morphologically, a
phase separation comparison suggests alignment
of the main X-ray peak with the main radio one, contrary to the ASCA finding 
by Takahashi et al (2001) which claim an
alignment with the radio interpulse. An accepted Rossi-XTE observation
will clarify this point.

We measure a soft power law spectral index ($\alpha \simeq 1.7$),
different from those measured for PSR J0218+4232 and PSR B1821$-$24.
The $N_{\rm H} \sim 2.3\times 10^{22}$ cm$^{-2}$ is marginally consistent
with the Galactic value in the direction of the pulsar and gives a
e$^-$/$N_{\rm H}$ ratio of 1/100, 10 times smaller than the canonical
average value. This could be justified by the position of \P\ in the
Galaxy w.r.t. our line of sight but could also suggest a larger
distance than the DM derived one.
Radio timing should settle this issue soon.

\vskip 0.4cm
\begin{acknowledgements}
This research is supported by the Italian Space Agency (ASI) and
Consiglio Nazionale delle Ricerche (CNR). BeppoSAX is a major program
of ASI with participation of the Netherlands Agency for Aerospace
Programs (NIVR).
\end{acknowledgements}
   
